\documentclass[12pt,a4paper]{article}

\usepackage{graphicx}
\usepackage{amssymb}
\usepackage{epstopdf}
\usepackage{amssymb,amsthm,amsmath,mathrsfs}

\newcommand{\B}{{\mathbf B}}
\newcommand{\M}{{\mathbf M}}
\newcommand{\Ss}{{\mathbf S}}
\newcommand{\T}{{\mathbf T}}

\renewcommand{\u}{{\mathbf u}}

\newcommand{\e}{{\mathbf e}}
\newcommand{\R}{\mathbb{R}}
\newcommand{\Z}{\mathbb{Z}}

\newtheorem{theo}{Theorem}

\newtheorem{cor}[theo]{Corollary}
\newtheorem{lem}[theo]{Lemma}

\title{Maslov Indices and Monodromy}
\author{HR Dullin \\ 
  Department of Mathematical Sciences \\ 
  Loughborough University \\
  Loughborough LE11 3TU \\
  {\small  H.R.Dullin@lboro.ac.uk} 
 \and
JM Robbins and  H Waalkens\\
School of Mathematics\\
University of Bristol\\
University Walk\\
Bristol BS8 1TW
 \and
SC Creagh and G Tanner \\
School of Mathematical Sciences\\
University of Nottingham\\
University Park\\
Nottingham NG7 2RD
 }
\date{April 2005}

\begin{document}

\maketitle

\abstract{
We prove that for a Hamiltonian system on a cotangent bundle
that is Liouville-integrable and has monodromy the vector of Maslov indices 
is an eigenvector of the monodromy matrix with eigenvalue 1.
As a corollary the resulting restrictions on the monodromy matrix are 
derived.
}

\section{Introduction}

The Liouville-Arnold theorem describes the local structure of an integrable 
system: for regular values of the energy-momentum map $F: T^*M \to
\R^n$, 
the preimage of 
a regular value 
is an $n$-torus (or a union of disconnected $n$-tori, but for
simplicity we assume there is just one),
and there exist action-angle
variables in a neighbourhood of this torus. 
Thus locally phase space has 
the structure of a trivial 
$n$-torus-bundle 
over an open neighbourhood of a regular
value in the image of $F$. 
Duistermaat \cite{Duistermaat80} pointed out that 
globally the torus-bundle
over the regular values of $F$ may be non-trivial. 
This phenomenon is called monodromy.
As a result there may not exist
global action-angle variables. 
In two degrees of freedom monodromy is well understood \cite{Matveev96,Zung97}.
It is a common phenomenon because it occurs in a neighbourhood of an 
equilibrium of focus-focus type. In three degrees of freedom now also many examples 
\cite{WD02,WDR03,DGC03,BDV05} are known.

Quantisation of a classical system with monodromy leads to quantum monodromy 
\cite{CushDuist88,VuNgoc99,Child98,SadovskiiZhilinskii99,ECS04}. 
The fact that the classical actions cannot be globally
defined implies that the quantum numbers suffer the same problem. 

The Maslov index is not only interesting for 
semiclassical quantisation, but also in classical mechanics it is an
invariant object defined for paths on Lagrangian submanifolds, e.g.\ 
on invariant tori, see \cite{Arnold67,Maslov81,AG80}.  
Recently it has been shown that the Maslov index is related to the singular points of the
energy-momentum map \cite{FoxmanRobbins05}.

In this letter we are going to show that if the vector of Maslov indices is non-zero, 
then it is an eigenvector of the monodromy matrix with eigenvalue 1.
This has some interesting consequences for the structure of admissible monodromy 
matrices.
Since the Maslov index is only defined on cotangent bundles our results
are only valid when the phase space is a symplectic manifold of the form $T^*M$.

\section{Maslov indices}


Let $C$ be a closed curve in the set of regular values of the
energy-momentum map. We take $C$ to be parameterised by $0 \le s
\le 1$.  Let $T_s$ denote the corresponding one-parameter family of
$n$-tori in phase space.  Fix a basis of cycles $\gamma_0$ for $T_0$.
By continuation this defines a basis of cycles $\gamma_s$ for every
$s$. The curve $C$ has monodromy when $\gamma_1 = \M \gamma_0$ for
$\M \in SL(n, \Z)$ is nontrivial.  More precisely, monodromy is a
nontrivial automorphism of the first homology group, and it implies
that the preimage of $C$ under $F$ is a nontrivial $n$-torus-bundle
over $C$.  The basis $\gamma_s$ determines actions $I_s$ and Maslov
indices $\mu_s$ on $T_s$.  In fact, the Maslov indices are
independent of $s$, as they depend continuously on $s$ and are
integer-valued \cite{Trofimov95}.  Let us denote their common value by
$\mu$.  Our main result is the following simple observation:

\begin{theo}\label{thm:maslov-indices}
If the vector of Maslov indices $\mu$ is not equal to zero, 
then $\mu$ is an eigenvector of the monodromy matrix $\M$ with eigenvalue 1.
\end{theo}

\begin{proof}
  
  We have that $\mu_1 = \M \mu_0$ (just as $I_1 = \M I_0$), since in
  general a change of basis cycles $\gamma' = \T
  \gamma$, where $\T \in SL(n,\Z)$, induces the transformation of
  Maslov indices $\mu' =
  \T \mu$ (and the transformation of actions $I' = \T I$).  Since $\mu_s =
  \mu$ for all $s$,
$\mu_1 = \M \mu_0$, i.e. 
\[
     \M \mu = \mu \,.
\]
\end{proof}
We remark that the  Maslov indices $\mu$, the actions $I$, and the
monodromy matrix $M$ depend on the initial
 choice of basis $\gamma_0$.  Under a change of basis $\gamma_0' =
 \T\gamma_0$, where $T \in SL(n,\Z)$, we have that $\mu' = \T\mu$ and $\M' = \T\M\T^{-1}$.

\section{Monodromy matrices}

From Theorem~\ref{thm:maslov-indices} we immediately obtain the well-known result \cite{Matveev96,Zung97} about the 
structure of monodromy matrices in two degrees of freedom:
\begin{cor}\label{cor:2}
For $n=2$ degrees of freedom and a loop $C$ with $\mu \not = 0$ there exists a basis of 
cycles such that the monodromy matrix of $C$ has the form 
\[
\M = \begin{pmatrix}
1 & m \\ 0 & 1
\end{pmatrix}
\].
\end{cor}
\begin{proof}
Since $\M \in SL(2,\Z)$ the eigenvalues $\lambda_1, \lambda_2$ must satisfy
$\lambda_1 \lambda_2 = 1$. But one eigenvalue must be 1 by Theorem~\ref{thm:maslov-indices}, 
hence $\lambda_1 = \lambda_2 = 1$. Finally a matrix in $SL(2,\Z)$ 
with a single eigenvalue equal to 1 is conjugate to the stated form by some matrix 
from $SL(2, \Z)$. The Maslov index in this basis is $\mu = (\mu_1, 0)$.
\end{proof}


Notice that this does not give a complete classification of monodromy matrices on 
cotangent bundles because we have assumed that $\mu \not = 0$. 
When $\mu \not = 0$, Corollary~\ref{cor:2} 
is quite strong because no assumption is needed on the type of singularity 
that is encircled by $C$, in particular the usual non-degeneracy condition is
not needed.

Corollary~\ref{cor:2} is a special case of the simple general
\begin{lem}\label{lem:monodromy-matrices}
Suppose $\M \in SL(n,\Z)$ has eigenvalue $\pm 1$.  Then there exists $\T \in
 SL(n,\Z)$ such $\M' = \T\M\T^{-1}$ has first column equal to $\pm\e_1 =
 (\pm 1, 0,\dots, 0)^t$.
\end{lem}
\begin{proof}
  Let $\u$ denote an eigenvector of $\M$ with eigenvalue $\pm 1$,
  chosen so that its components are coprime integers. Then one can
  construct a matrix $\Ss \in SL(n,\Z)$ whose first column is $\u$
  (see, e.g., \cite{Cassels59}).  Let $\T = \Ss^{-1}$ and $\M' =
  \T\M\T^{-1}$.  It is easy to check that $\e_1$ is an eigenvector of
  $\M'$ with eigenvalue $\pm 1$, so that $\M'$ has first column equal
  to $\pm \e_1$.
\end{proof}

Using Lemma \ref{lem:monodromy-matrices} and again the fact that $\det \M = 1$ and $\lambda_1 = 1$,
we can obtain the classification of monodromy matrices (for non-zero Maslov index)
in $n=3$ degrees of freedom:
\begin{cor}\label{cor:4}
For $n=3$ degrees of freedom and a loop $C$ with $\mu \not = 0$ there exists a basis of 
cycles such that the monodromy matrix $\M$ of $C$ has one of the following forms: 
\begin{equation}
  \label{eq:1}
\begin{pmatrix}
1 & * & *  \\
 0 & 1 & * \\
 0 & 0 & 1 
\end{pmatrix}, \quad
 \begin{pmatrix}
1 & * & *  \\
 0 & -1 & * \\
 0 & 0 & -1 
\end{pmatrix}, \quad
\begin{pmatrix}
1 & * \\
0 & \B 
\end{pmatrix}, \quad
\end{equation}
where $\B \in SL(2, \Z)$ has irrational eigenvalues and $*$ denotes integers.
\end{cor}
\begin{proof}
The eigenvalue 1 can appear with algebraic multiplicity $m_a = 1$ or $m_a= 3$ only; 
$m_a = 2$ is  impossible because $\det \M = \lambda_1 \lambda_2 \lambda_3 = 1$. 
\footnote{For general $n$ the multiplicity cannot be $n-1$.}
The case $m_a = 3$ corresponds to the first form above.
When $m_a = 1$, the remaining eigenvalues are either both $-1$,
corresponding to the second form,
or they are irrational, corresponding to the third. 
Other combinations of eigenvalues are not possible, because
rational eigenvalues of matrices in $SL(n,\Z)$ are necessarily equal
to $\pm 1$.  

If the eigenvalues are all $\pm 1$ (corresponding to the first two
forms), the matrices can be made upper triangular by applying 
Lemma \ref{lem:monodromy-matrices} recursively, using a transformation of the form
\[
  \T_n = \begin {pmatrix}
1 & * \\
{\bf 0} & \T_{n-1}
\end{pmatrix} \,.
\]
Matrices with two irrational eigenvalues
cannot be made upper triangular in $SL(n,\Z)$
(as the diagonal elements of a triangular matrix are its eigenvalues). 
\end{proof}

It is interesting to consider how the entries denoted $*$ in
(\ref{eq:1}) can be normalised.  In the first form the eigenvalue $1$
has geometric multiplicity $m_g$ equal to $1$ or $2$ (i.e., there are
either one or two independent eigenvectors with eigenvalue $1$).  The normal form
for $m_g = 2$ has been computed in \cite{WD02}.  The result is that
only a single nonzero element remains above the diagonal. Essentially
this means that when $m_g= 2$ the matrix can be block-diagonalised in
$SL(3,\Z)$.  In the remaining cases in (\ref{eq:1}) a block-diagonal
form is in general not possible:
Conjugating a block triangular matrix with a block triangular matrix gives
\[
\begin{pmatrix}
1 & -{\bf d} {\bf D}^{-1} \\
{\bf 0} & {\bf D}^{-1}
\end{pmatrix}
\begin{pmatrix}
1 & \bf{a} \\
{\bf 0} & {\bf A}
\end{pmatrix}
\begin{pmatrix}
1 & {\bf d}  \\
{\bf 0} & {\bf D}
\end{pmatrix}
=
\begin{pmatrix}
{\bf 1} & ( {\bf a} - {\bf d} {\bf D}^{-1} ({\bf A} - {\bf 1})) {\bf D} \\
{\bf 0} & {\bf D}^{-1}{\bf A}{\bf D}
\end{pmatrix} \,.
\]
Setting the upper right element of the right hand side 
to zero and solving for ${\bf d}$ involves the inverse of ${\bf A - 1}$ 
which is in general not an integer matrix. Using a more general transformation
leads to the same condition.
Thus for general ${\bf a}$ and ${\bf A}$ the monodromy
matrix cannot be block-diagonalised. However, e.g., for the special matrix 
${\bf A} = \begin{pmatrix} 2 & 1 \\ 1 & 1\end{pmatrix}$ (the `cat map')
it is always possible since $\det( {\bf A - 1} ) = -1$. If ${\bf A - 1}$ is 
singular (corresponding to the first case in Corollary 4) the resulting
Diophantine equation may or may not have a solution.

Our results need the condition $\mu \ne 0$. It may be possible to show that $\mu \ne 0$ 
necessarily holds for certain configuration spaces $M$.
We suspect, for example, that this is the case when $M=\R^n$, 
although we have not been able to prove this.
Our result would then give the complete classification of monodromy matrices on $T^*\R^n$. 
In particular the construction of arbitrary monodromy matrices given in 
\cite{CushmanVuNgoc02} would be impossible on these cotangent bundles.

In three degrees of freedom, the known examples of monodromy are
either of the first form with $m_g = 2$  \cite{WD02, WDR03} 
or of the last form and block-diagonal. 
The last form of $\M$ is realised for geodesic flows
on $Sol$-manifolds, where an arbitrary hyperbolic $\B \in SL(2, \Z)$
may appear \cite{BDV05}.

The main implication of the above is that when $m_a = 3$ and $m_g = 2$ 
there are always two invariant actions, i.e.\ actions that do not change globally
along the path $C$. 
Obviously there is always one invariant action, namely the one corresponding
to the eigenvector $\e_1$, and when $m_g = 1$ it is the only one.
With eigenvalues $-1$ there is at most one invariant action, but another 
action is invariant when $C$ is traversed twice. 
Hence on a covering space this may reduce to $m_a = 3$ and $m_g = 2$.
It would be very interesting to find an example of this type.


\section*{ Acknowledgement }
HRD would like to that AP Veselov and VS Matveev for helpful discussions. This work was 
supported by the EU Research Training Network ''Mechanics and Symmetry in 
Europe'' (MASIE), (HPRN\_CT-2000-00113).

\bibliographystyle{plain}

\def\cprime{$'$} \def\cprime{$'$} \def\cprime{$'$}

\end{document}